\documentclass[aps,twocolumn,groupedaddress]{revtex4}
\usepackage{graphicx}
\usepackage{amsmath,amssymb, fixmath}

\begin{document}

\newcommand{\etal}      {{\it et~al.}}


\title{Impact of geometry on the magnetic flux trapping of superconducting accelerating cavities }



\author{D. Longuevergne$^1$ and A. Miyazaki$^2$}
\affiliation{$^1$Université Paris-Saclay, CNRS/IN2P3, IJCLab, France, $^2$FREIA Laboratory, Uppsala University, Sweden}



\begin{abstract}
Controlling trapped magnetic flux in superconducting radiofrequency (RF) cavities is of crucial importance in modern accelerator projects.
In order to study flux trapping efficiency and sensitivity of surface resistance,
dedicated experiments have been carried out on different types of low-$\beta$ superconducting accelerating cavities.
Even under almost full trapping conditions, we found that the measured magnetic sensitivities of these cavity geometries were significantly lower than the theoretical values predicted by commonly-used models based on local material properties. 
This must be resolved by taking account of geometrical effects of flux trapping and flux oscillation under RF surface current in such cavity shape.
In this paper, we propose a new approach to convolute the influence of geometries. 
We point out a puzzling contradiction between sample measurements and recent cavity experiments, 
which leads to two different hypotheses to simulate oscillating flux trapped in the cavity surface.
A critical reconsideration of flux oscillation by the RF Lorentz force, compared with temperature mapping studies in elliptical cavities, favoured the results of previous sample measurements, which suggested preferential flux trapping of normal component to the cavity inner surface.
Based on this observation, we builded a new model to our experimental results and the discrepancy between old theory and data were resolved.
\end{abstract}

\maketitle

\section{Introduction\label{sec:intro}}
Performances of a superconducting accelerating cavity made of bulk niobium can be strongly affected by the presence of a residual magnetic field while transitioning into the superconducting state. 
Magnetic vortices trapped by pinning centers of the cavity material during cool-down interact strongly with the radiofrequency (RF) electromagnetic fields and induce additional dissipations in the helium bath~\cite{padamsee1998}. 
We denote these additional losses by a surface resistance $R_{\rm mag}$ in n${\rm \Omega}$.

The total surface resistance $R{\rm s}$ of a superconductor~\cite{Gurevich17} is the sum of a strongly temperature $T$ dependent contribution $R_{\rm BCS}$, 
derived from the linear response of the Bardeen Cooper Schrieffer theory of superconductivity~\cite{mattis58}, 
and the other contribution, which is only weakly temperature dependent, defined as the residual resistance $R_{\rm res}$: 
\begin{equation}
R_{\rm s} (f, T) = R_{\rm BCS}(f, T) + R_{\rm res} (f, T) \label{eq:surface_decompose},
\end{equation}
with RF frequency $f$ dependence.
This $R_{\rm res}$ consists of a component temperature-independent $R_0$ due to material imperfections (pollution, defects, grain boundaries, etc) and $R_{\rm mag}$:
\begin{equation}
R_{\rm res} (f, T) = R_0 + R_{\rm mag} (f, T) \label{eq:residual_decompose}.
\end{equation}

Recent technical advances on mechanical process and surface treatment have reduced $R_0$, 
and therefore, understanding and controlling $R_{\rm mag}$ becomes of critical importance in state-of-the-art superconducting RF cavities for many applications~\cite{padamsee2009}. 
Previous studies~\cite{doi:10.1063/1.4903808, doi:10.1063/1.4927519, PhysRevApplied.5.044019, posen16, PhysRevAccelBeams.22.032001} mainly focused on the specific elliptical cavities dedicated to high-energy electron accelerators. 
In this paper, we report on three more general shaped cavities developed for proton and heavy ion linear accelerators. 
The first is a Quarter-Wave Resonator (QWR) operating at a frequency of 88 MHz, built for the Spiral2 project~\cite{Marchand:2015gcn}. 
The second is a Double-Spoke Resonator (DSR) operating at 352 MHz for the ESS project~\cite{duchesne2013}. 
The third is a Single-Spoke Resonator (SSR) operating also at 352 MHz for the MYRRHA project~\cite{Longuevergne:2017kfe}. 

With the ambient residual magnetic field $H_{\rm res}$ in mG present at transition, $R_{\rm mag}$ can be decomposed into:
\begin{equation}
R_{\rm mag} = \eta_{\rm mag} \cdot S_{\rm mag}(f, T) \cdot H_{\rm res}, \label{eq:Rmag_decompose}
\end{equation}
where $\eta_{\rm mag}$ is the dimensionless flux trrapping efficiency coefficient with $0<\eta_{\rm mag}<1$~\cite{benvenuti1997}, $S_{\rm mag}$ is the magnetic sensitivity~\cite{doi:10.1063/1.5016525} expressed in n${\rm \Omega}$/mG.

From Eq.~(\ref{eq:Rmag_decompose}), three independent factors, $H_{\rm res}$, $\eta_{\rm mag}$, and $S_{\rm mag}$ play an important role to reduce $R_{\rm mag}$ and thus to fulfill the requirement for total $R_{\rm s}$ of each accelerator project. 
The purpose of this study is to address geometrical dependence of $R_{\rm mag}$ when a condition $\eta_{\rm mag}\sim1$ is satisfied. 
We show that the origin of such geometrical dependence may be from a preferential flux trapping angle in the cavity surface.
This influences $S_{\rm mag}$ that is mainly contributed from flux trapped to normal to the surface.

This report is organized as follows. 
In the rest of the introduction, we review the previous studies of $H_{\rm res}$, $\eta_{mag}$, and $S_{\rm mag}$. 
In~\ref{sec:experiment}, the experimental set-up at Ir\`{e}ne Joliot Curie Laboratory (IJCLab) is presented. 
Section~\ref{sec:expulsion} is dedicated to the experimental results on flux trapping and the diamagnetic effect.
In section~\ref{sec:sensitivity}, experimental results on flux sensitivities on our cavities are described.
In order to explain our experimental findings, section~\ref{sec:discussion} discusses several aspects of flux trapping and sensitivities.
Comparing previous sample experiments, we point out one puzzle in recent cavity experiments.
We critically reconsider recent theories and propose a hypothesis about geometrical dependence of $R_{\rm mag}$, which is justified by recent studies on single-cell elliptical cavities tested in unusual configuration with a horizontal magnetic field and slow cooling down..
We then compare our experimental results and this new model.
The final section represents conclusions.

\subsection{Magnetic shield to reduce $H_{\rm res}$}
To protect niobium from environmental magnetic field coming from the earth field and magnetic parts at the vicinity of a cavity,
magnetic shields are usually installed around a superconducting cavity. 
High permeability material as permalloy ($\mu$-metal), Cryophi\textregistered or A4K are used to funnel the magnetic field and thus strongly attenuate $H_{\rm res}$ inside it. 
Historically, minimizing the attenuation factor, required to achieve $R_{\rm mag}$ within the total budget of $R_{\rm s}$, has been a major interest of the community~\cite{crawford2014study}, 
and this has been successful for conventional elliptical cavities at 1.3~GHz. 

One technical difficulty appears when one tries to shield cavities with a different geometry, namely, low-$\beta$ structures~\cite{Facco_2016}. 
Their relatively large dimensions can significantly increase the mechanical complexity and cost to fabricate ideal magnetic shields~\cite{Zheng:2017edf} and thus can practically limit field attenuations around the cavities. 
Therefore, reducing the other two factors in Eq.~(\ref{eq:Rmag_decompose}) becomes motivated to relax the mechanical and financial constraint of magnetic shielding. 
On top of this practical use, systematic studies of $\eta_{\rm mag}$, and $S_{\rm mag}$ are of scientific interest for the applied superconductivity under strong RF fields~\cite{benvenuti99}.

\subsection{Flux expulsion and trapping}
The flux trapping efficiency coefficient $\eta_{\rm mag}$ is typically evaluated with magnetic sensors installed in close proximity with a cavity~\cite{PhysRevAccelBeams.22.032001}. 
The sensor probes the magnetic field distribution altered by the diamagnetic property of the Meissner state when the material goes through superconducting transition. 

Amongst all, bulk material history has a significant impact on flux trapping. 
Material re-crystallization by thermal treatment typically above 800~$^\circ$C would ensure an almost complete flux expulsion beside some exceptions as reported [8]. 
Indeed, without any recrystallization process, close to 100\% of the residual magnetic field is trapped as reported in~\cite{benvenuti1997, PhysRevSTAB.15.062001, vallet1994}. 
Regarding the studies presented in this paper, the cavities are made of polycrystalline material without heat treatment above 650~$^\circ$C and therefore we can primarily assume, based on past observations, almost full flux trapping.

The previous studies about cool-down dynamics sometimes showed contradictory results. 
For simple geometries like bare elliptical cavities in vertical cryostats~\cite{doi:10.1063/1.4903808, posen16, doi:10.1063/1.4875655}, 
the flux expulsion improves in proportion to the thermal gradient across the cavity. 
This was explained by two different models proposed by Kubo~\cite{10.1093/ptep/ptw049} and Checchin~\cite{PhysRevApplied.5.044019}. 
However, for more general configurations, a low thermal gradient is proposed in Ref.~\cite{PhysRevSTAB.16.102002} due to the possible effect of the thermoelectric currents generated by bi-metallic junctions between the niobium cavity and the helium tank made of titanium~\cite{PhysRevSTAB.18.042001, PhysRevAccelBeams.22.052001}.

The experiments on the LCLS-II cryomodules~\cite{wu_SRF2017, wu2019achievement} showed that such dynamic thermoelectric currents tend to vanish when the temperature is close to superconducting transition. 
We also observed a similar behaviour in our cryostat at IJCLab~\cite{Longuevergne:2015ass} and thus we do not address the effect of thermoelectric current in this paper. 
In LCLS-II, independently of the cool-down rate, the intrinsic static thermoelectric currents remain and act as an additional external magnetic field to be expelled during the superconducting transition. 
It must be noted that the cavities installed in other cryostats or cryomodules may behave differently due to the different mechanical structure from our vertical test stand.

\subsection{Magnetic sensitivity $S_{\rm mag}$}
A static model concerning a normal conducting core in a trapped vortex gives a good approximation of $S_{\rm mag}$ as formulated by~\cite{padamsee1998}
\begin{equation}
S_{\rm mag} = \frac{R_{\rm n}(f, T)}{2\cdot H_{\rm c2}(T)}, \label{eq:old_model}
\end{equation}
with $R_{\rm n}$ the normal resistance and $H_{\rm c2}$ the upper critical field of the material.
As described in Appendix~\ref{sec:flux_oscillation_GR}, dynamic flux oscillation under Lorentz force driven by RF current leads to the same result, based on the model by Gittleman and Rosenblum~\cite{gittleman65}. 
Beyond this simple approximation, the magnetic sensitivity is extremely difficult to quantitatively predict and evaluate since it depends on many parameters~\cite{PhysRevAccelBeams.23.023102}: 
\begin{itemize}
\item	Frequency of the cavity~\cite{doi:10.1063/1.5016525, vallet1994}
\item	Temperature of operation~\cite{ono1999, Longuevergne:2019sfi}
\item	Local heating due to trapped vortices~\cite{PhysRevB.77.104501, gurevich13}
\item	Impurity content of material: dislocations, segregation, precipitates, and grain boundaries~\cite{checchin17}
\item	Model of pinning potential~\cite{vaglio18}
\item	Interplay of various pinning centers~\cite{liarte18, PhysRevApplied.14.044018}
\item	Amplitude of RF fields~\cite{benvenuti99, vaglio18, Longuevergne:2019sfi, liarte18, PhysRevApplied.14.044018, PhysRevAccelBeams.22.073101}
\item	Geometry of the cavity~\cite{benvenuti99, Longuevergne:2019sfi, liarte18}
\end{itemize}

One must take care when studying the geometrical dependence, 
because the measurement observable $R_{\rm mag}$ averaged over the cavity inner surface cannot directly separate the geometrical effects from either $\eta_{\rm mag}$ or $S_{\rm mag}$. 
To study $S_{\rm mag}$, one needs to ensure $\eta_{\rm mag}\sim 1$ by very low thermal gradients generally associated with slow cooling speed. 
Our study fulfills this condition because the material is prepared {\it not} to efficiently expel the flux during cool-down. 
We show this by a dedicated experiment in section~\ref{sec:expulsion}.

The geometrical dependence of $S_{\rm mag}$ has been under debate. 
A group reported~\cite{Kramer:2019nzd} that flux trapping happens uniformly and homogeneously over the surface and even preserves the orientation of the applied external field before superconducting transition, 
if flux expulsion is suppressed by slow cooling down. 
The flux oscillation under RF fields depends on the orientation of the trapped flux versus RF currents and thus becomes non-uniform over the surface. 
Although this experimental result based on magnetic field mapping~\cite{Kramer:2019nzd} seems convincing and plausible,
we recognized a puzzling contradiction against rather general studies on superconducting samples in other experiments, such as Ref.~\cite{Candia_1999}.
Our experimental results suggest careful argumentation of this subtle issue.
In section~\ref{sec:discussion}, we come back to this point and critically reconsider existing theories and experiments.
We propose a hypothesis that flux trapping may happen preferentially for the normal component to the surface, 
and the amount of flux to be oscillated by the RF fields becomes non-uniform. 
This results in $S_{\rm mag}$ being dependent on the geometry.
Our hypothesis successfully explains the distribution of RF power dissipation measured in temperature mapping~\cite{Kramer:2019nzd}.

\section{Experimental setup\label{sec:experiment}} 
\subsection{The vertical cryostat}
\begin{figure}[h]
\includegraphics[width=50mm]{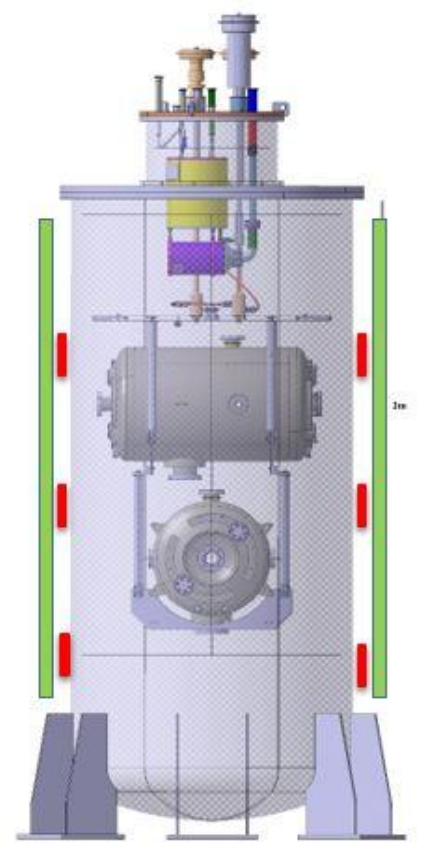}
\caption{
Vertical cryostat in operation hosted in SUPRATECH facilities at IJCLab laboratory (Orsay, France) with its passive and active magnetic shields. The insert is loaded with two ESS Double Spoke Resonators. SUPRATECH facilities are dedicated to surface processing and testing of SRF cavities.
\label{fig:vertical}}
\end{figure}

In an effort to fully qualify any new cavity design or a new surface treatment process or procedure, 
cavities are first tested in a vertical cryostat. 
The vertical cryostats are designed to provide optimal testing conditions to address cavity performances. 
The intrinsic quality factor $Q_0$ is evaluated at different accelerating gradients $E_{\rm acc}$ by measuring the power dissipation $P_{\rm c}$ averaged over the cavity walls. 
The cryostat, available on platform Supratech at IJCLab in operation since 1998 and upgraded in 2018, 
is capable of hosting two cavities (equipped with their helium jacket) in a volume constrained in a cylinder of 2~m high and 1.15~m in diameter as shown in Fig.~\ref{fig:vertical}.

\begin{figure}[h]
\includegraphics[width=80mm]{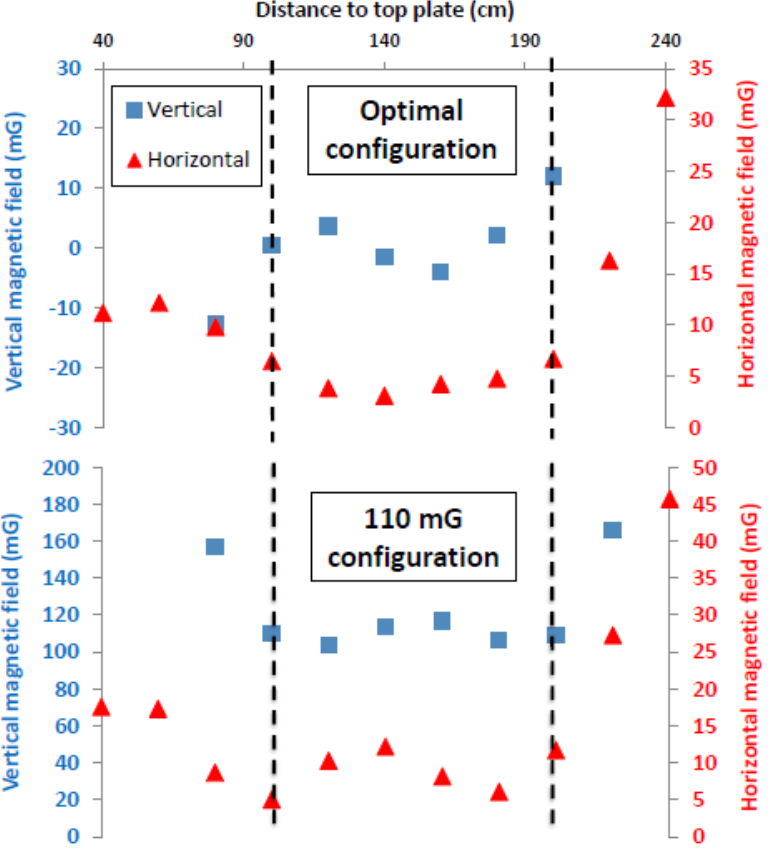}
\caption{
Example of residual magnetic field configuration applied in the vertical cryostat. The vertical and horizontal components are measured along the central axis of the cryostat.
\label{fig:B_field}}
\end{figure}

The cryostat is externally shielded on the side by 1~mm-thick permalloy sheets rolled around the vacuum vessel. 
The horizontal component of the magnetic field is significantly attenuated. 
Because of design constraints, the vertical component is not shielded by permalloy sheets but by three compensating coils inserted in between the magnetic shield and the vacuum vessel as depicted in Fig.~\ref{fig:vertical}. 
This configuration allows the possibility to either reduce the magnetic field to a minimum value to optimize the cavity performances or to apply a uniform field to measure $\eta_{\rm mag}$ and/or evaluate $S_{\rm mag}$ with the field of any cavity. 
Figure~\ref{fig:B_field} shows two examples of magnetic configurations of a residual field.

\subsection{Magnetic sensors}
Regarding magnetic measurement, the very low magnetic field to be measured makes the fluxgate magnetometer the best technology in our test conditions in vacuum at low temperatures. 
As commercially available fluxgate sensors (at the time of these studies) are only available as single axis sensors, 
three of them are assembled on a 3D-printed support to measure each of the three axis. 
A home-made multiplexer has been built to read up to twelve type G sensors with only one controller (MAG01-H) from Bartington~\cite{Bartington}. 
The magnetic field resolution is 0.02~mG over a range of 20~G. 
Integration time imposes a minimum multiplexing rate of about six seconds. 
To avoid any crosstalk between sensors, both current and voltage leads are multiplexed. 
In normal operations, only one sensor is energized at a time.

\subsection{Measurement capabilities}
Several types of measurements are possible with the current set-up:
\begin{itemize}
\item Evaluation of the magnetic shield efficiency
\item Evaluation of the flux trapping efficiency $\eta_{\rm mag}$~\endnote{this is usually not precise as the cavities are equipped with a helium tank, which limits the accessibility for instrumentation.}
\item Evaluation of the magnetic sensitivity $S_{\rm mag}$ to a residual magnetic field of different types of superconducting cavities 
\item Monitoring of the magnetic field behaviour during cooling down generated by thermoelectric currents because of the existence of bi-metallic junctions
\item Detection of the magnetic field penetration in case of a quench
\end{itemize}

\subsection{Cavities for this study}
\begin{figure}[h]
\includegraphics[width=85mm]{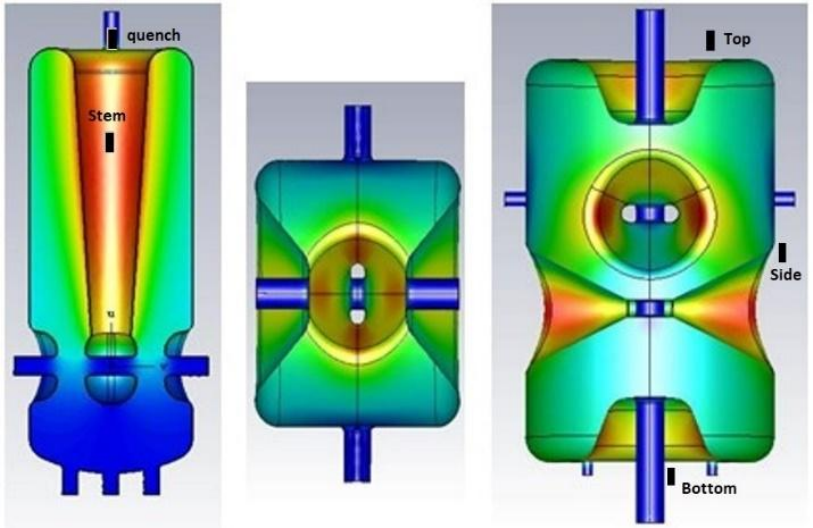}
\caption{
From left to right, RF magnetic field distribution of: Spiral2 QWR, MYRRHA Single Spoke Resonator (SSR) and ESS Double Spoke Resonator (DSR). Black rectangles represent the position of magnetic field probes installed during experiments.
\label{fig:cavities}}
\end{figure}

The three cavities dedicated for this study are shown in Fig.~\ref{fig:cavities} with geometrical factor $G$~\cite{padamsee1998} and operating frequency $f$ in Table~\ref{tab:cavities}. 
The cavities are made of polycrystalline niobium and their surfaces have been prepared following the standard procedure at IJCLab laboratory:
\begin{itemize}
\item Degreasing in an ultrasonic bath with detergent.
\item Surface abrasion by Buffered Chemical Polishing of at least 200~um (BCP)
\item Optional hydrogen degassing at 650~$^\circ$C for 10~h.
\item High Pressure Rinsing with ultra-pure water
\item Drying and assembly in ISO4 clean room
\end{itemize} 
They are all manufactured out of the same polycrystalline (fine grain) bulk niobium material without any heat treatments above 800~$^\circ$C.
No low temperature baking~\cite{padamsee2009} or nitrogen doping~\cite{grassellino13} were performed. 
In addition, all cavities for this experiment are equipped with a titanium helium tank.

Vertical and transverse residual magnetic fields were applied on Spiral2 QWR.
For MYRRHA SSR and ESS DSR, only the vertical fields were applied in the cryostat. 
This corresponds to the field transverse to the beam axis in MYRRHA and along the beam axis of ESS DSR. 
At the time of these studies, the ESS DSR could only be loaded vertically in the cryostat as shown in Fig.~\ref{fig:cavities}. 
The field orientations are also summarized in Table~\ref{tab:cavities}.

\begin{table*}[bth]
  \centering
  \begin{tabular}{ccccc} \hline
    Type of cavity      & project & $G$ ($\Omega$) &  $f_0$ (MHz) & Applied magnetic field orientation \\ \hline
    QWR                 & SPITAL2 & 33             &  88          & vertical and transverse \\
    SSR                 & MYRRHA  & 109            &  352         & vertical (transverse to the beam axis) \\
    DSR                 & ESS     & 133            &  352         & vertical (along the beam axis) \\ \hline
  \end{tabular}
  \caption{Geometrical factors and frequencies of the cavities in this experiment}
  \label{tab:cavities}
\end{table*}

\section{Flux trapping study\label{sec:expulsion}}
We first validate our assumption about almost full flux trapping in our fine grain material without substantial heat treatment above 650~$^\circ$C. 
The conventional experiment on $\eta_{\rm mag}$ by fluxgate sensors around a cavity is not reliable in our experimental setup because of the complicated cavity geometries studied in this paper compared to the conventional elliptical cavities.
The maximum field enhancement even by the ideal flux expulsion is only a few percent around the cavity. 
Instead, a magnetic sensor is installed to monitor the vertical component of the magnetic field {\it in the stem}, 
the inner conductor of the Spiral2 QWR as depicted in Fig.~\ref{fig:cavities}. 
A particular advantage of using QWR for this study is that the magnetic sensor can be installed inside the cavity structure, 
unlike the conventional elliptical cavities, in which the beam vacuum side is not suitable for sensor installation.

\begin{figure}[h]
\includegraphics[width=85mm]{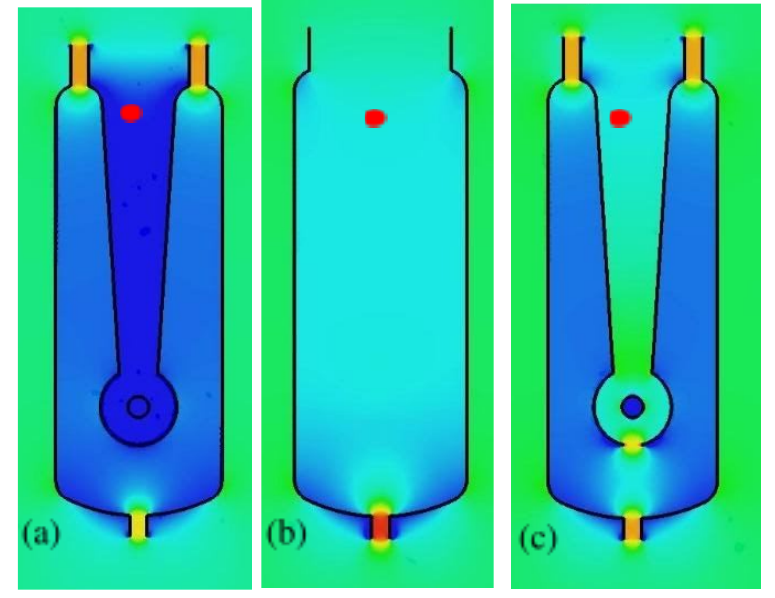}
\caption{
Vertical component of the magnetic flux distribution calculated with CST Studio Suite considering (a) full expulsion, 
(b) full expulsion with the inner conductor normal conducting 
and (c) full expulsion with trapping at the bottom of the stem. 
The magnetic fields measured at the probe (red dot) are respectively 0\%, 38\% and 48\% of the ambient residual field. 
The ambient vertical field is vertical.
\label{fig:QWR_expulsion}}
\end{figure}

In case of complete flux trapping during superconducting transition, 
no change of magnetic field would be observed whereas in the case of complete flux expulsion, 
the magnetic field would drop to zero as shown in Fig.~\ref{fig:QWR_expulsion}(a). 
As an intermediate case, Fig.~\ref{fig:QWR_expulsion}((b) shows the partial diamagnetic effect by the outer conductor when the inner conductor is still normal conducting.

The particular case of complete flux expulsion with the concentrated flux trapped at the bottom of the stem is also considered as shown in Fig.~\ref{fig:QWR_expulsion}(c). 
Indeed, during regular cooling down, the SC/NC interface moves from the bottom of the outer conductor, then to the top of the cavity and finally down to the bottom of the inner conductor. 
This would result in a non-zero field measured in the inner conductor even with complete flux expulsion. 
The magnetic sensor installed inside the inner conductor (see red dots in Fig.~\ref{fig:QWR_expulsion}) allows to distinguish between different types of trapping scenario in Fig.~\ref{fig:QWR_expulsion}. 

\begin{figure}[h]
\includegraphics[width=85mm]{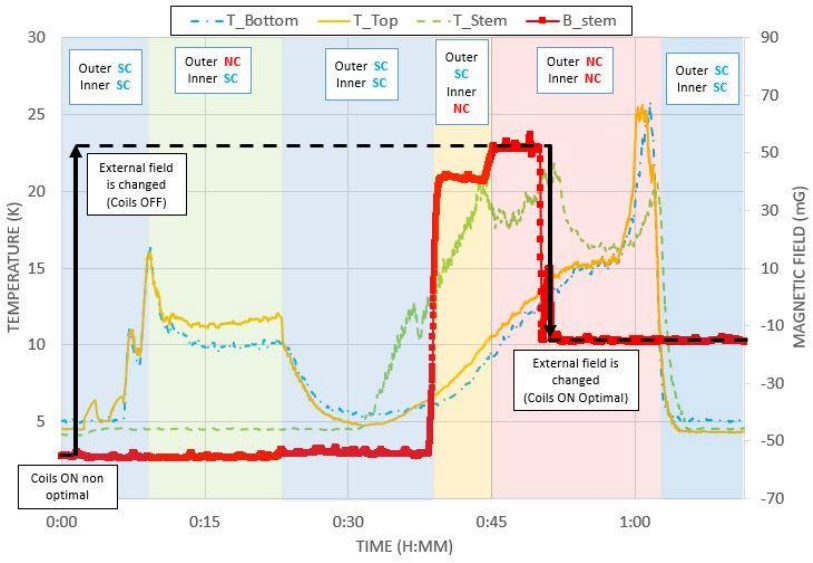}
\caption{
Flux trapping experience on a Spiral2 QWR.
\label{fig:QWR_expulsion_hist}}
\end{figure}

The flux trapping experiment has been performed as follows with the experimental data summarized in Fig.~\ref{fig:QWR_expulsion_hist}
\begin{enumerate}
\item The cavity has been cooled down to 4~K in a non-optimal ambient magnetic field, resulting in a residual field measured by the probe of $-55$~mG shown as a blue region noted Outer SC and Inner SC. 
\item The ambient magnetic field is changed (Helmholtz coils are off) to generate a magnetic field of$+52$~mG at t = 0:00 indicated by a black arrow still in the blue region. No reaction of the magnetic probe inside the stem proves the perfect diamagnetic behaviour (Meissner effect) at 4~K.
\item The external conductor is warmed up with heaters above transition shown as the light green region noted Outer NC and Inner SC. No reaction of the magnetic probe indicates complete shielding by the inner conductor (stem).
\item The outer conductor is cooled below transition again. Some amount of flux could be trapped during transition of the outer conductor but did not affect the sensor inside the stem because of perfect shielding by the inner conductor.
\item The inner conductor (stem) is warmed up with a heater above transition shown as the orange region noted Outer SC and Inner NC. The magnetic field measured by the probe changes significantly to reach $+41$~mG, indicating inefficient Meissner shielding of the outer conductor.
\item The outer conductor is warmed up above transition. At this stage, the entire cavity is normal conducting shown as the red region noted Outer NC and Inner NC. The magnetic field finally reaches $+52$~mG, which is the true ambient field without any field distortion by the Meissner effect.
\item The compensating coils are set to optimize ambient magnetic field ($-15$~mG).
\item The cavity is then cooled down below transition with a cooling rate of 112~mK/s and a temperature gradient of 4 K between top and bottom of the stem, shown as a blue region noted Outer SC and Inner SC. No reaction of the magnetic probe was observed. This indicates full flux trapping in both inner and outer conductors.
\end{enumerate}

From Fig.~\ref{fig:QWR_expulsion}(b), whatever the flux expulsion efficiency during cool down is, 
the outer conductor acts as only a weak magnetic shield at 4~K for the inner conductor. 
This is due to the presence of the three coupler ports at the bottom of the cavity and 2 ports on the top. 
Therefore, even in case of perfect diamagnetic effect by the material, the outer conductor would shield 62\% of the ambient field. 
However, in reality, because of the poor flux expulsion by the polycrystalline niobium, the shielding is only about 10\% (the difference of magnetic field measured between step 5 and 6). 
The shielding capability of the outer conductor could thus not explain the very low $R_{\rm mag}$ of this cavity to the vertical magnetic field, 
and this needs other reasons as discussed from the next section.
Note that no finite flux jump was observed during the cool-down. 
This indicates that the flux trapping is uniform and not like Fig.~\ref{fig:QWR_expulsion}(c). 
A related subject concerning quench induced trapped flux is summarized in the Appendix~\ref{sec:quench}.

From this experimental results, the following statements can be concluded. 
The inner conductor provides a very efficient magnetic shielding at 4~K, 
whereas, even at 4 K, the simulation shows that the outer conductor has a very weak shielding capability due to the presence of ports at the top and bottom of the cavity. 
Flux trapping during cooling down is close to 100\% in agreement with our assumption of the fine grain material. 
Even the inner conductor does not act as an efficient shield against permanent ambient fields present during superconducting transition.
However, the inner conductor can very efficiently shield fields activated during accelerator operation from solenoids. 

This experiment helps us to understand the global flux trapping in our low-$\beta$ structure.
No flux expulsion due to thermal gradient was observed and the fluxes are almost fully trapped during cooling down.
However, this global flux trapping is an observation of a bulk property of a cavity material with a magnetic sensor in a certain distance.
Any surface processes regarding flux exlulsion might not be resolved by this method.

\section{Magnetic sensitivity\label{sec:sensitivity}}
From now on, we assume that the flux is almost fully trapped in the surface of the cavity and no global flux expulsion happens in the cavities under test. 
At this stage, we do not assume any angles of trapped flux at the inner surface of the cavities.
For the precise experiment of $S_{\rm mag}$, one has to remove the components $R_{\rm BCS}$ and $R_0$ to extract pure $R_{\rm mag}$. 
This is accomplished by the following procedure:
\begin{enumerate}
\item The cavity is slowly cooled down in an ambient residual magnetic field $H_0$ as low as possible, the so-called “optimal configuration” depicted previously in Fig.~\ref{fig:B_field}. The vertical component stays below 10~mG as well as the horizontal component
\item The total surface resistance $R_{\rm s}$ at low field is estimated from the $Q_0$ measurement
\begin{equation}
R_{\rm s} = \frac{G}{Q_0},
\end{equation}
with $G$ the geometrical factor of the cavity as listed in Table~\ref{tab:cavities}.
\item The cavity is warmed up slowly above transition during a night ($>50$~K) and then cooled down in a homogeneous vertical magnetic field $H_1 = 110$~mG. The horizontal component stays below 15~mG.
\item $R_{\rm s}$ is measured again, considering that the flux is fully trapped. By subtracting two surface resistances, we can estimate $S_{\rm mag}$ by 
\begin{equation}
S_{\rm mag} = \frac{R_{\rm s} (H_1) - R_{\rm s} (H_0)}{H_1 - H_0}.
\end{equation}
\end{enumerate}

The sensitivities measured on the three types of cavities are summarized in Table~\ref{tab:comp_result}. 
The simple model calculation based on Eq.~(\ref{eq:old_model}) overestimates the measured sensitivities of all the cavities. 
Apparently, recently proposed approaches with flux oscillation~\cite{vaglio18, gurevich13, checchin17, liarte18, PhysRevApplied.14.044018} based on the Bardeen-Stephen model~\cite{bardeen65} would improve the calculation 
but lack of precise information on material parameters especially on pinning centers prevents us from applying their models. 
Besides, the intrinsically complicated shape of the cavities would certainly limit the accuracy of these models, 
which were originally developed and validated for simpler geometries, such as 1-cell elliptical cavities. 
In the next section, we develop a novel way to take account of geometrical effects while keeping the local surface resistance the same as Eq.~(\ref{eq:old_model}). 
As discussed in Appendix~\ref{sec:flux_oscillation_GR}, Eq.~(\ref{eq:old_model}) can be considered as a special case of a local surface resistance based on the classical Lorentz force model by ignoring material dependence in detail.
The simple convolution of geometry, which we introduce in the next section, leads to dramatically better agreements with all the measurement results as shown in Table~\ref{tab:comp_result}.

\begin{table*}[tbh]
  \centering
  \begin{tabular}{ccccccc} \hline
    Type of cavity      & \begin{tabular}{c} $H_{\rm res}$            \\  orientation                       \end{tabular} 
                        & \begin{tabular}{c} measurement              \\  (${\rm m\Omega/mG}$)                    \end{tabular} 
                        & \begin{tabular}{c} uniform $S_{\rm mag}$    \\  (${\rm m\Omega/mG}$) Eq.~\ref{eq:old_model} \end{tabular} 
                        & \begin{tabular}{c} relative                 \\  error \%                          \end{tabular}
                        & \begin{tabular}{c} corrected $S'_{\rm mag}$ \\  (${\rm m\Omega/mG}$) Eq.~\ref{eq:new_model} \end{tabular} 
                        & \begin{tabular}{c} relative                 \\  error \%                          \end{tabular}  \\ \hline

    QWR                 & vertical   & 0.006         &  0.08    & $+93$ & 0.011 & $+45$ \\
    QWR                 & horizontal & 0.05          &  0.08    & $+38$ & 0.048 & $-4$ \\
    SSR                 & vertical   & 0.043         &  0.12    & $+64$ & 0.047 & $+8.5$ \\
    DSR                 & beam axis  & 0.06          &  0.12    & $+50$ & 0.055 & $-9$ \\ \hline
  \end{tabular}
  \caption{Comparison of measured and calculated sensitivities}
  \label{tab:comp_result}
\end{table*}

\section{Discussions\label{sec:discussion}}
\subsection{Geometrical dependence and trapped flux angle}
In Table~\ref{tab:comp_result}, the measured sensitivities are systematically and significantly lower than the calculated sensitivities based on the model 
which assumes uniform power dissipation over the inner surface for all the geometries. 
Moreover, the difference in sensitivity of the QWR by a factor of 10 to a vertical or horizontal field suggests a very strong geometrical dependence. 

The fact that RF magnetic fields are mainly distributed around the inner conductor, where the surface is almost vertical, 
could explain these observations. 
Indeed, as the surface resistance is estimated from the power dissipations, 
a change in surface resistance could be measured if and only if it occurs in high RF magnetic field regions. 
Trapping flux on the bottom of the QWR where the RF electric fields dominate, for example, 
does not induce any $Q_0$ drop as only RF magnetic fields are dissipating. 

The $R_{\rm mag}$ is no longer uniform over the cavity surface but shows position dependence. 
In this description, the angle between applied field $H_{\rm res}$ and the surface {\it does} play a major role on the local surface resistance. 
However, there exist two hypotheses to explain this angular dependence.
\begin{description}
\item [Case 1] At the surface, flux perpendicular to the surface is more preferentially trapped and thus drives the RF losses.  
\item [Case 2] The flux components are homogeneously trapped and the flux oscillation by RF depends on the angle between the trapped flux and the surface.
\end{description}
At the first glimpse, they seem not distinguishable by RF measurement.
In this paper, we show that we can in fact distinguish them by temperature mapping.

The case 2 is suggested by recent measurements using magnetic field mapping around an elliptical cavity~\cite{Kramer:2019nzd}.
In this scenario, the orientation of the flux is preserved before and after the superconducting phase transition,
if global flux expulsion is suppressed by uniform cooling down especially on polycrystalline material without annealing over 800$^\circ$C.
Whatever the trapped angle is, the trapped flux is vibrated by RF current via the Lorentz force.

\subsection{Previous sample measurement suggests case 1}
We point out that the case 1 is strongly favoured by previous theoretical and experimental studies.
Experiments~\cite{Candia_1999, PhysRevB.56.2809} on the reversible magnetization $\mathbold{M}_{\rm eq}$ and irreversible magnetization  $\mathbold{M}_{\rm irr}$ verified the angular dependence on trapped flux applied on isotropic type-II thin films. 
The magnetization is a macroscopic measure of trapped vortices' orientation. 
In equilibrium magnetization with applied field $H$ higher than the lower critical field $H_{\rm c1}$, $\mathbold{M}_{\rm eq}$ and correspondingly vortex lines are normal to the surface at low fields and become aligned with an externally applied field $\mathbold{H}$ when it approaches the upper  critical field $H_{\rm c2}$. 
On the other hand, the pinning effect $\mathbold{M}_{\rm irr}$ is always normal to the surface regardless of the field strength if the angle between the applied field and the normal vector to the surface is within certain value determined by the geometry (smaller than 70 degree in their samples). 

One may argue that the dimension of the sample measurement is so thin that these results are not relevant to be compared with cavity materials.
However, near $T_{\rm c}$, the penetration depth is infinitely long so that the cavity wall can be treated as a thin film. 
The cavity cool-down virtually mimics the magnetization measurement of a small sample 
because cooling under a constant small external field is the same as reducing the external field at constant temperature.
In addition, another experiment on a sample for cavities showed that the flux trapping is more likely in the normal direction to the surface~\cite{Eichhorn:2017rme}.
This experiment was based on fluxgate sensors.

The fundamental reason of perpendicular flux trapping can be explained by the image force effect.
The image force which acts on a parallel component of trapped flux is so strong that any realistic pinning force in relatively clean niobium can {\it not} keep the flux within a few penetration depths.
This effect is only at the surface,
 and the macroscopic measurement by magnetic sensors, as described in section~\ref{sec:expulsion} may not be capable to resolve it.
The detail discussion is summarized in Appendix~\ref{sec:image_force}.

\subsection{Lorentz force}
The sample measurement and direct measurement of elliptical cavities with fluxgate sensors are clearly in contradiction.
In order to distinguish case 1 and 2, we consider Lorentz force caused by RF current at the surface.
Here, we clarify different spatial distribution of RF losses in case 1 and case 2.
The contributions of $S_{\rm mag}$ based on flux oscillation can be estimated by considering the Lorentz force interaction between the quantized flux line and the local RF current. 

\subsubsection{Case 2: no angle dependence in flux trapping}
\begin{figure}[h]
\includegraphics[width=85mm]{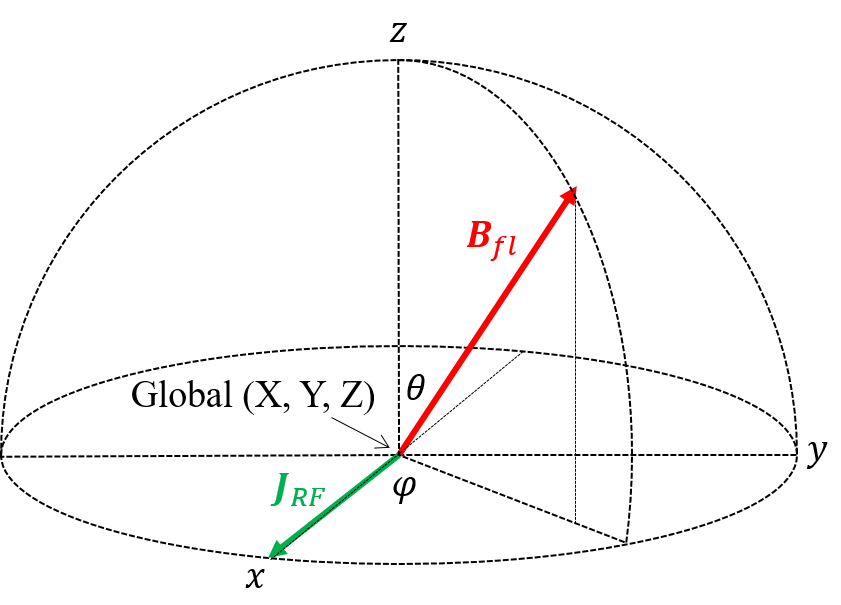}
\caption{
Local spherical coordinate system with RF current aligned in the x-axis and trapped flux pointing $(\theta, \phi)$.
\label{fig:coordinate}}
\end{figure}

We start to consider case 2 first.
As shown in Fig.~\ref{fig:coordinate}, with a local spherical coordinate with an RF current density vector $\mathbold{J}_{\rm RF}$ aligned to the x-axis, 
$\theta$ the polar angle between the RF surface and the trapped flux vector $\mathbold{B}_{\rm fl}$, 
and $\varphi$ the angle between the trapped flux projected to the RF surface and the RF current, 
the Lorentz force density vector $\mathbold{f}_{\rm L}$ can be written as
\begin{eqnarray}
\mathbold{f}_{\rm L} & = & \mathbold{J}_{\rm RF} \times \mathbold{B}_{\rm fl} \nonumber
\end{eqnarray}
\begin{eqnarray}
=
\left(
\begin{array}{c}
J_0 \\
0   \\
0   \\
\end{array}
\right)
\times 
\left(
\begin{array}{c}
\phi_0\cos{\varphi}\sin{\theta} \\
\phi_0\sin{\varphi}\sin{\theta} \\
\phi_0\cos{\theta} \\
\end{array}
\right)
=
\left(
\begin{array}{c}
0 \\
-J_0 \phi_0\cos{\theta} \\
J_0 \phi_0\sin{\varphi}\sin{\theta} \\
\end{array}
\right)
\end{eqnarray}
where $\phi_0$ is the flux quantum ($2.07\times10^{15}$~Wb) and $J_0$ is the amplitude of the RF current density.
In this model, $\left(\theta, \varphi \right)$ is determined by the relative orientations of the magnetic residual field, 
the cavity surface and the direction of the RF currents in the particular position.

To calculate flux oscillation in a thick bulk niobium in y-direction, Checchin et al.~\cite{checchin17} introduced a 1-dimensional differential equation 
\begin{equation}
M\frac{d^2y}{dt^2} + \eta \frac{dy}{dt} + f_{\rm p} = -J_0 \phi_0 \cos{\theta}
\end{equation}
with $M$ and $\eta$ the effective mass and viscosity of a flux, respectively, in the Bardeen-Stephen model and $f_{\rm p}$ the pinning force~\endnote{Note that Ref.~\cite{checchin17}takes $\theta\rightarrow\pi/2-\theta$ in their coordinate system.}.
This equation originates from Gittleman and Rosenblum~\cite{gittleman65},
who calculated flux oscillation in a thin film 12.7~um thick, and thus $\theta\sim0$ was fairly applied in their case.
However, in case 2 for general shaped cavities, $\theta\sim\pi/2$ can happen and this corresponds to a flux trapped in the parallel direction to the surface.
Therefore, there may be another degree of freedom (z-direction) in the flux oscillation
\begin{equation}
M\frac{d^2z}{dt^2} + \eta \frac{dz}{dt} + f_{\rm p} = J_0 \phi_0 \sin{\varphi} \sin{\theta}. \label{eq:Lorentz_parallel}
\end{equation}

Providing that the material is uniform and isotropic, $M$, $\eta$ and $f_{\rm p}$ are similar in both y- and z-axis, 
and these two degrees of freedom contribute to two independent modes of the flux oscillation. 
Correspondingly, the sum of these two modes results in power dissipation. 
In both flux flow and pinning regimes, magnetic sensitivity is locally
\begin{equation}
S_{\rm mag} \propto \cos^2{\theta} + \sin^2{\theta}\sin^2{\varphi}. \label{eq:S_case2}
\end{equation}

\subsubsection{Case 1: perpendicular trapping to the surface}
In the first order approximation of case 1, only the perpendicular component of the flux is trapped and contributes to the Lorentz force
\begin{eqnarray}
\mathbold{f}_{\rm L} & = & \mathbold{J}_{\rm RF} \times \mathbold{B}_{\rm fl} \nonumber
\end{eqnarray}
\begin{eqnarray}
=
\left(
\begin{array}{c}
J_0 \\
0   \\
0   \\
\end{array}
\right)
\times 
\left(
\begin{array}{c}
0 \\
0 \\
\phi_0\cos{\theta} \\
\end{array}
\right)
=
\left(
\begin{array}{c}
0 \\
-J_0 \phi_0\cos{\theta} \\
0 \\
\end{array}
\right)
\end{eqnarray}
The flux oscillation becomes virtually 1-dimensional and mathematically reproduces the formulation by Ref.~\cite{checchin17}.
The magnetic sensitivity is locally
\begin{equation}
S_{\rm mag} \propto \cos^2{\theta}.\label{eq:S_case1} 
\end{equation}
The predicted angular dependence is different from the former model Eq.~(\ref{eq:S_case2}), in which all the flux components are trapped whichever the orientation to the RF surface is.
This difference can be experimentally addressed in the method that we propose in the next section.

\begin{figure*}[tb]
\includegraphics[width=85mm]{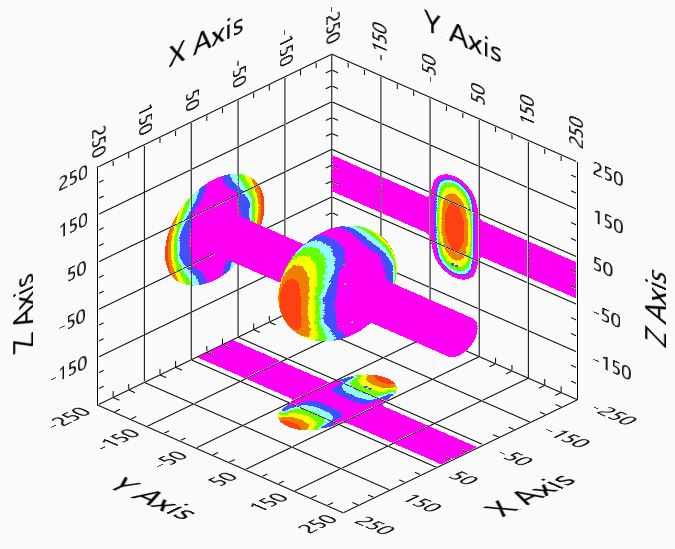}
\includegraphics[width=85mm]{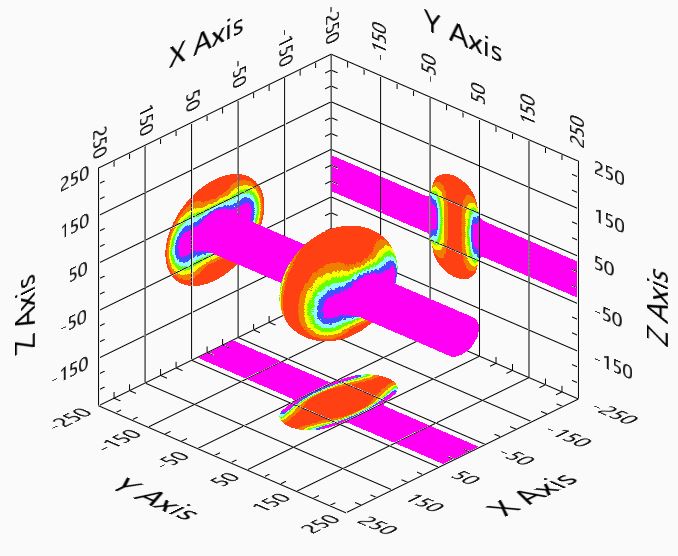}
\caption{
Comparison of RF power dissipations predicted by case 1 (left) and case 2 (right). The external magnetic field is applied along the x-axis. This field is perpendicular to the beam axis of the elliptical cavity.
\label{fig:elliptical_case1_case2}}
\end{figure*} 

\subsection{Convolution of geometrical effect and local sensitivity}
We developed a numerical method to distinguish case 1 from case 2 by convoluting Eq.~(\ref{eq:S_case2}) or Eq.~(\ref{eq:S_case1}) and cavity geometry with Eq.~(\ref{eq:old_model}).
The local spherical coordinate $(\theta, \varphi)$ in the previous discussion is defined at each point over the cavity surface in the global Cartesian coordinate $(X, Y, Z)$, 
and namely, the Lorentz force and $S_{\rm mag}$ depend on $\theta(X, Y, Z)$ and $\varphi(X, Y, Z)$. 
When we apply $\mathbold{H}_{\rm res}$ to a cavity, $\theta(X, Y, Z)$ and $\varphi(X, Y, Z)$ are determined by the three dimensional structure of the cavity. 
We take account of this effect as follows:
\begin{enumerate}
\item Numerically evaluate the residual field to be trapped 
\begin{equation}
H_\perp (X, Y, Z) = \frac{\left| \mathbold{J}_{\rm RF}\times \mathbold{H}_{\rm res}\right|}{J_0}
\end{equation}
from three dimensional models of cavities. 
In case 1, $H_\perp (X, Y, Z)$ becomes $H_{\rm res}\cos{\theta}(X, Y, Z)$.
\item Evaluate the local surface resistance caused by the trapped flux oscillation
\begin{equation}
R_{\rm mag}(X, Y, Z) = R_{\rm n} (f, T) \frac{H_{\perp} (X, Y, Z)}{2H_{\rm c2}}.\label{eq:local_R_mag}
\end{equation}
from Eq.~(\ref{eq:Rmag_decompose}) and Eq.~(\ref{eq:old_model}) with $\eta_{\rm mag}\sim 1$ as discussed before. 
It is important to emphasize that this formulation is the same as the consequence of Lorentz force calculation in Gittleman and Rosenblum~\cite{gittleman65} as described in Appendix~\ref{sec:flux_oscillation_GR}.
The material dependence is included in $R_{\rm n}(f, T)$ and $H_{\rm c2}$, and fixed in this analysis which focuses on geometrical effects.
\item Similar to the previous work by one of the authors~\cite{LONGUEVERGNE201841}, integrate Eq.~(\ref{eq:local_R_mag}) all over the cavity surface $S$ and obtain a new sensitivity $S'_{\rm mag}$ with geometrical correction
\begin{equation}
S'_{\rm mag} = \frac{\int_S R_{\rm mag}(X, Y, Z) H^2_{\rm RF}(X, Y, Z) dS}{ H_{\rm res}\int_S H^2_{\rm RF}(X, Y, Z) dS} \label{eq:new_model}
\end{equation}
with $H_{\rm RF}$ the local RF magnetic field evaluated in the same model as for $H_\perp$.
\end{enumerate}
This model is computed in a NI LabVIEW software~\cite{LabVIEW} with exported files generated by CST Microwave Studio, such as  and  distributions and surface mesh~\cite{CST}. 

\subsection{Comparison to an elliptical cavity}
Figure~\ref{fig:elliptical_case1_case2} shows RF power dissipations ($\propto R_{\rm mag}(X, Y, Z)$) from our model applied to an elliptical cavity when the external magnetic field is perpendicular to the beam axis.
The configuration of the magnetic fields corresponds a typical horizontal cavity test.
The case 1 shows two dissipating areas aligned on the equator and placed at 180$^{\circ}$ from each other.
On the other hand, case 2 shows rather uniform dissipation along on the equator.
This is because the angle $\varphi$ is approximately 90$^\circ$ between the flux perpendicularly applied to the cavity and the RF current around the equator so that
\begin{equation}
S_{\rm mag} \propto \cos^2{\theta} + \sin^2{\theta} \sim 1
\end{equation}
is independent of the polar angle $\theta$.

Temperature mapping data showed in Figure 12 in Ref.~\cite{Kramer:2019nzd} clearly highlights that dissipating regions draw two domains centred on the equator where the applied field is normal to the surface and separated of 180$^\circ$. 
No increase of surface resistance is measured where trapped flux is parallel to the surface. 
This observation is in favour of the case 1, in which flux in the surface is trapped in the perpendicular direction.
The parallel components, which should exist in case 2,  may not contribute to the RF dissipation or may be simply expelled from the surface due to the image force.

It is very important to note that vertical tests, in which $\varphi\sim0$ is applied around the equator, are not suitable to distinguish case 1 from case 2 because the Lorentz force depends only on $\cos{\theta}$ in both case 1 and case 2.
The different dependence on $\theta$ and $\varphi$ in case 1 and 2 is hidden in vertical tests.
However, the horizontal tests can reveal the difference between case 1 and 2.

In order to investigate our model further, 
the same calculations of case 1 are applied to different angles between the external magnetic field and the beam axis of the elliptical cavity.
We compare our geometrically corrected $S'_{\rm mag}$ and the experimental data taken by the same group of temperature mapping~\cite{Kramer:2019nzd}. 
Their experiment was conducted in a similar configuration as ours. 
They deliberately applied 100~mG to a cavity with different angles, and several thermal cycles ensured $\eta_{\rm mag}$ higher than $0.5$ and even close to $0.9$. 
Note that this group concluded homogeneous trapping without any angular preference, contrary to our above argument.
Their conclusion was based on the magnetic field mapping outside the cavity. 

\begin{figure}[h]
\includegraphics[width=85mm]{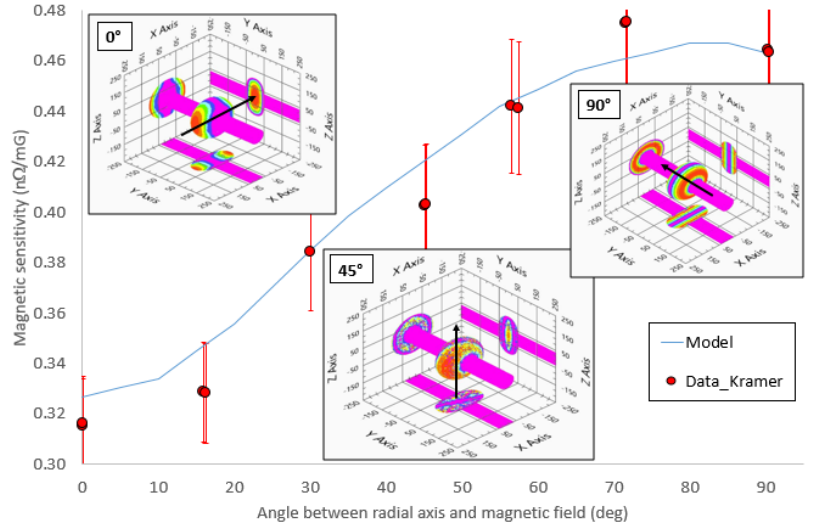}
\caption{
Angular dependence of the magnetic sensitivity for an elliptical cavity. The model (blue line) predicts the magnetic sensitivities measured in [40] (Red dots). 
The sensitivities calculated by the model is scaled by a factor of 3.6 to fit experimental values within 5\% error.
\label{fig:comp_elliptical}}
\end{figure}

We remove a common residual resistance $R_0=5\, {\rm n}{\rm \Omega}$~\cite{Kramer_private} from published $R_{\rm res}$~\cite{Kramer:2019nzd} and divide it by 100~mG to obtain $S_{\rm mag}$. 
The absolute value of $S_{\rm mag}$ depends on material parameters and is not predictable.
For the correction of the material effect, we linearly scale the model prediction $S'_{\rm mag}$ to fit the data. 
As depicted in Fig.~\ref{fig:comp_elliptical}, quantitative agreement is obtained between our model and their experimental data. 
The RF power dissipation is more sensitive to higher angle closer to the magnetic field parallel to the beam axis. 
Although the RF field between the iris and the equator is lower than that at the equator, 
the total area of trapped flux normal to the cavity surface is large so that Eq.~(\ref{eq:new_model}) leads to higher sensitivity.

It has to be pointed out that experimental data published in Ref.~\cite{doi:10.1063/1.4927519} showed on the contrary that the transverse sensitivity ($=0^\circ$) is higher than the axial sensitivity ($=90^\circ$). 
This difference can be explained, on one hand, by the capability of the material to expel the magnetic flux instead of fully trapping it, and on the other hand, by the cooling configuration. 
Indeed, the expelled vortices are pushed by the SC/NC interface and can be concentrated on the last remaining normal conducting region.
In Ref.~\cite{doi:10.1063/1.4927519}, the cavity is set horizontally and cooled from the bottom to the top. 
The last remaining normal region where all vortices are concentrated is at the top of the equator, the most sensitive region. 
They achieved an excellent flux expulsion by heat treatment above 800$^\circ$C with nitrogen doping and also by a fast cool-down. 
On contrary, the study by Ref.~\cite{Kramer:2019nzd} was dedicated for almost full flux trapping and global flux expulsion was suppressed. 
Trapped flux is thus more uniformly spread over cavity surface and not only concentrated in very sensitive regions as the equator. 
Therefore, the conclusions can be different in each case.
The material for our coaxial cavities is not prepared for efficient flux expulsion and our experimental condition is closed to Ref.~\cite{Kramer:2019nzd}.

\subsection{Application to our cavities}
\begin{figure}[h]
\includegraphics[width=85mm]{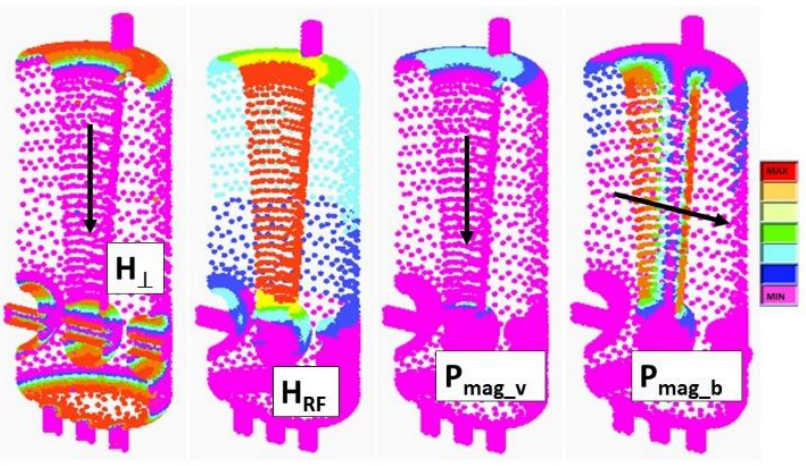}
\caption{
Outputs from LabVIEW routine showing from left to right: the normal trapped magnetic field ($H_\perp$) under a vertical $H_{\rm res}$, the RF field distribution ($H_{\rm RF}$), and the normalized power dissipations caused by trapped flux from vertical $H_{\rm res}$ ($P_{\rm mag_v}$) and horizontal $H_{\rm res}$ ($P_{\rm mag_b}$) for the Spiral2 QWR.
\label{fig:model_QWR}}
\end{figure}

Figure~\ref{fig:model_QWR} depicts the three kinds of graphical output generated by the code.
The results of this model $S'_{\rm mag}$ are shown in Table~\ref{tab:comp_result}. 
The relative errors between the experimental and calculated values with the proposed model are dramatically improved by the proposed geometrical correction. 
This good agreement is consistent with the analysis on elliptical cavities and consolidates case 1.
The flux in the coaxial-type cavities is almost fully trapped but the amount of the flux normal to the surface, which contributes to the RF power dissipation, depends on the relative angle between the applied field and the cavity surface. 

\subsection{Puzzle from the magnetic field measurement}
Our model, motivated by sample measurements, agrees on RF measurement results and temperature mapping data.
This apparently contradicts with magnetic field mapping~\cite{Kramer:2019nzd} and even our measurement by a fluxgate sensor in section~\ref{sec:expulsion}.
We do not have direct evidence to fully explain this puzzling contradiction.
We just propose one hypothesis which could resolve the issue.
It is well known that global flux expulsion is influenced by bulk property of the material.
The surface property has been considered to just affect the flux sensitivity in the first approximation.
However, we argue that the surface property could be particularly important to the expulsion of parallel flux within RF penetration depth.
Such a phenomenon is so local at the inner surface of the cavity that magnetic field sensors in certain distance from the surface may not resolve very small distortion in the parallel field.

\begin{figure}[h]
\includegraphics[width=85mm]{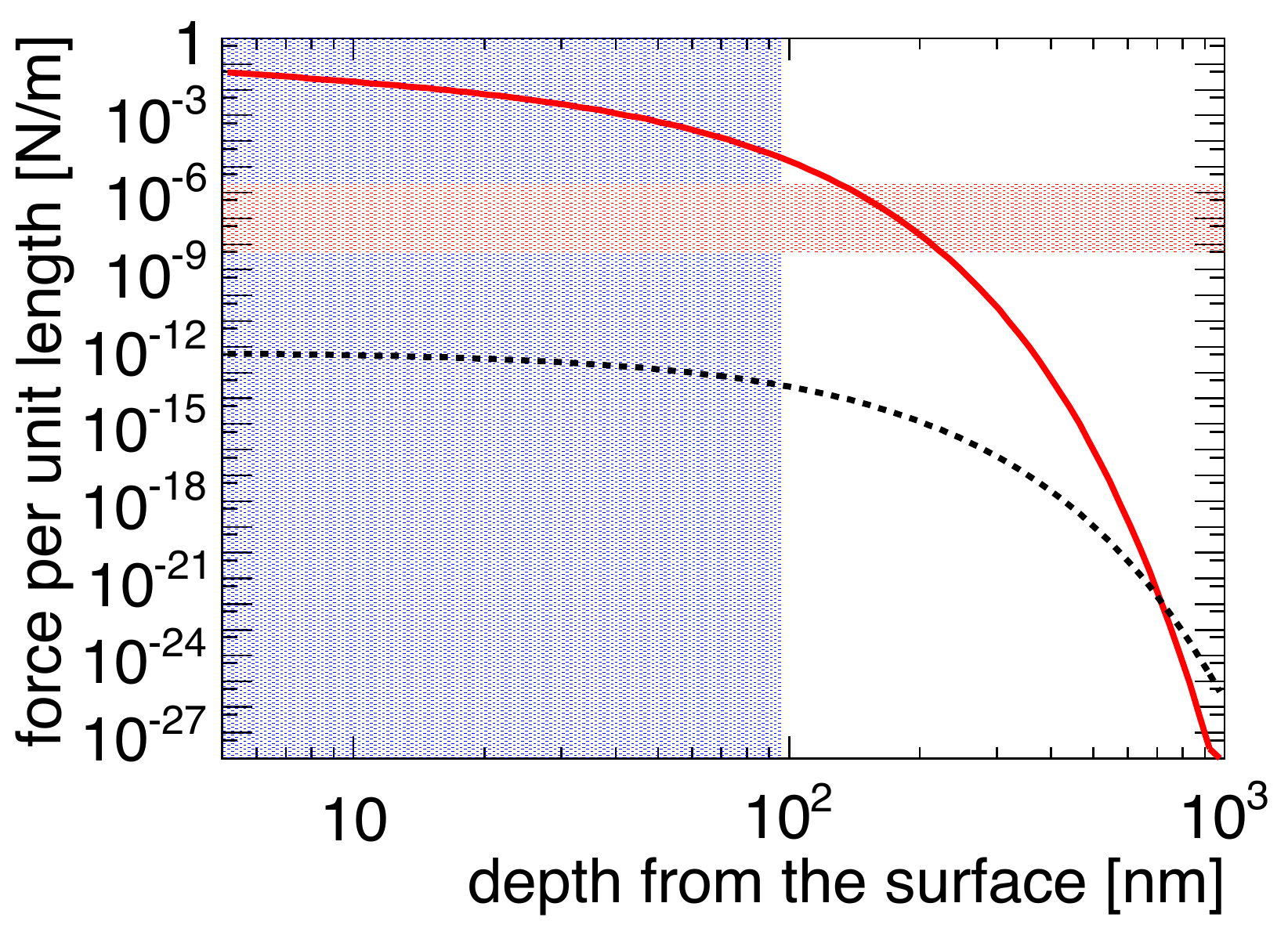}
\caption{
Comparison of image force $f_1$ (red solid line) external field interaction $f_2$  (black dashed line), pinning force $f_{\rm p}$  (horizontal red hatch) and $3\lambda$ RF penetration region (vertical blue hatch)
\label{fig:image_force}}
\end{figure}

As described in Appendix~\ref{sec:image_force}, three forces are important to understand the expulsion of a parallel flux.
The first is an image force $f_1$ which expells the flux, 
the second is a counter-acting force $f_2$,
and the third is pinning force $f_p$.
Figure~\ref{fig:image_force} compares the above three forces $f_1$, $f_2$, and $f_{\rm p}$, estimated from plausible parameters of our cavities, with a region of RF field penetration by $3\times \lambda$. 
Since the external field is as small as 100~mG, 
$f_2$ is also small so that the surface barrier of Bean and Livingston exists around 800~nm deep inside the bulk. 
Therefore, this effect does not keep parallel flux inside the RF penetrating region. 
Although the $f_{\rm p}$ estimate shows huge uncertainty, the image force is still stronger than the pinning force by several orders of magnitude.
This indicates that parallel fluxes, which can contribute to the RF power dissipation, are totally expelled from the surface. 

From this consideration, we can argue that the observation in Ref.~\cite{Kramer:2019nzd} may not contradict our statement. 
The outer surface of a cavity is usually not as clean as the inner surface and thus the pinning force must be stronger than the above estimation. 
The flux at the equator may be trapped in parallel at the outer surface and does not contribute to the RF power dissipation. 
The magnetic sensor placed outside the cavity with some distance may not resolve the parallel flux at the inner surface. 
If the parallel fluxes were trapped in the inner surface within a few penetration depths, 
the other oscillation mode in Eq.~\ref{eq:Lorentz_parallel} would change the spatial distribution of power dissipation.

Finally, we stress that this discussion may be special for our cavities, which is made of relatively clean niobium at the inner surface. 
For cavities after low temperature baking or nitrogen doping, 
the impurity content just underneath the inner surface within penetration depth is known to be substantial~\cite{padamsee2009, doi:10.1063/1.4974909}.
This would enhance the surface barrier to protect the RF field penetrating into the bulk but the same barrier would prevent the parallel ambient flux escaping from the bulk. 
From this consideration, such cavities might keep some amount of parallel trapped flux, which may show additional heat dissipation in Eq. (10).

\section{Conclusion}
Magnetic sensitivity measurements were performed at IJCLab on several type of resonators (QWR, single and double spoke). 
Unlike during cooling down, once the cavity becomes superconducting, magnetic shielding is complete. 
A magnetic shield made of superconducting material is very efficient to shield any magnetic field absent during cool-down, such as fields generated by coils or solenoids for accelerator operation. 
However, a superconducting shield made of poorly expelling material, such as reactor grade niobium, is totally inefficient to shield any ambient magnetic field during cooling down.
The complete understanding of the magnetic flux trapping mechanism and the ensuing magnetic sensitivity is of first importance to optimize cooling procedures and magnetic shielding for future projects.

Our measurements reveal a strong geometrical dependence of surface resistance to magnetic field. 
The real sensitivity, evaluated indirectly and globally by RF power measurements, is consistently lower than the theoretical sensitivity under an assumption of uniformly dissipating trapped flux. 
This demanded a new method to convolute the geometrical effect to calculate proper magnetic sensitivity. 
We pointed out that recent experiments based on magnetic field sensors around the cavities may contradict with well-known sample experiments.
The former suggest homogenous flux trapping without any angle dependence between the external magnetic field and cavity surface.
The latter suggests more efficient trapping to the perpendicular direction to the surface.
We critically reconsidered the flux oscillation under the Lorentz force effect driven by RF current.
The assumption, that the normal component of the residual magnetic field is preferentially trapped at the inner surface during the superconducting transition, appears to be a reasonable hypothesis by comparing elliptical cavity results. 
More particularly, the model can explain the {\it hot zone} distribution of the temperature mapping and the angular dependence of magnetic sensitivity.
Using this new method, a very good agreement between calculated and measured sensitivities has been obtained for several types of geometries in this study. 
The magnetic sensitivity would be thus not only determined by the material history and its capacity to trap magnetic vortices but also by the material surface orientation versus the ambient magnetic field during superconducting transition. 
The ambient flux parallel to the surface would be locally expelled during superconducting transition by the image force effect over a depth around the London penetration depth at the inner surface of the cavities. 

\section*{Acknowledgement}
The authors would like to thank all the experimental works and efforts of Supratech team and the accelerator department at IJCLab to allow such specific studies to happen on prototypes cavities dedicated to projects. 
Our special thanks goes to R. E. Laxdal for the useful discussions.
Part of this study (MYRRHA SSR) happened during MYRTE project, which has received funding from the EUR-ATOM research and training programme 2014-2018 under grant agreement N$^\circ$662186. 

\appendix
\section{Relation between a static model and a dynamics model of flux oscillation\label{sec:flux_oscillation_GR}}
In Ref.~\cite{padamsee1998}, the formulate Eq.~(\ref{eq:old_model}) is derived from a static model regarding normal conducting cores of trapped flux.
This can be understood as a special case of dynamics flux oscillation driven by the Lorentx force.
We derive this by combining the notation of Ref.~\cite{checchin17} and Ref.~\cite{vaglio18}.
The model is based on Ref.~\cite{bardeen65} for originally a thin film sample~\cite{gittleman65}.
Although the model disregards some fundamental aspects of trapped fluxes, such as effective flux tension, 
it is known to sufficiently explain some important experimental results.

We take the local coordinate system as shown in Fig.~\ref{fig:coordinate} but we swap the polarity of $z$-axis for simplicity.
For a perpendicularly trapped flux, the equation of motion of a unit length of the flux at depth $z$ can be
\begin{equation}
M\frac{d^2y}{dt^2} + \eta \frac{dy}{dt} + ky = f_{\rm L} = \phi_0 J_0 e^{-z/\lambda}e^{i\omega t} \label{eq:flux_motion}
\end{equation}
where we consider that the pinning potential is harmonic $ky^2/2$ and the RF current is reduced by the penetration depth $\lambda$.
A particular solution of this driven damped harmonic oscillator can be obtained by substituting a trial solution
\begin{equation}
y(t) = (y_1 + iy_2)e^{i\omega t} \label{eq:trial_solution}
\end{equation}
to Eq.~(\ref{eq:flux_motion}), comparing real and imaginary part
\begin{eqnarray}
y_1 & = & \frac{\phi_0J_0e^{-z/\lambda}}{(k-M\omega^2)^2 + (\eta \omega)^2} (k-m\omega^2) \\
y_2 & = & \frac{\phi_0J_0e^{-z/\lambda}}{(k-M\omega^2)^2 + (\eta \omega)^2} (\eta\omega)
\end{eqnarray}
so that
\begin{equation}
\frac{dy}{dt} = \frac{\omega\phi_0J_0e^{-z/\lambda}}{(k-M\omega^2)^2 + (\eta \omega)^2} \left[-(\eta\omega) + i(k-M\omega^2)\right]e^{i\omega t}
\end{equation}

The RF power dissipation by work to a flux can be
\begin{equation}
P_{\rm flux} = \int_{0}^{\infty} dz \frac{1}{T}\int_{0}^{T} dt\mathrm{Re}\left(f_{\rm L}\right) \mathrm{Re}\left(\frac{dy}{dt}\right),
\end{equation}
where $T=2\pi/\omega$ is one period of oscillation. For $n$ fluxes per unit area, 
\begin{equation}
P_{\rm mag} = nP_{\rm flux} =\frac{1}{4} \frac{n\omega^2 \eta\phi_0^2 J_0^2 \lambda}{(k-M\omega^2)^2 + (\eta \omega)^2}.
\end{equation}
Using normal conducting resistivity ratio $\rho_{\rm n}$, the Bardeen-Stephen model defined
\begin{eqnarray}
\eta = \frac{\phi_0 B_{\rm c2}}{\rho_{\rm n}},
\end{eqnarray}
and we obtain
\begin{equation}
P_{\rm mag} = \frac{1}{2} \frac{\omega^2}{(\omega_0-M\omega^2/\eta)^2 + (\omega)^2} \frac{n\phi_0}{2B_{\rm c2}} \frac{\rho_{\rm n}}{\lambda} (J_0\lambda)^2,
\end{equation}
where we defined $\omega_0 = k/\eta$
Finally, we substitute following relations of normal conducting surface resistance $R_{\rm n}$, surface RF magnetic field $H_{\rm RF}$ and magnetic field density of trapped flux $B_{\rm res}$
\begin{eqnarray}
R_{\rm n} & = & \frac{\rho_{\rm n}}{\lambda} \\
H_{\rm RF} & = & J_0\lambda \\
B_{\rm res} & = & n \phi_0,
\end{eqnarray}
and obtain
\begin{eqnarray}
P_{\rm mag} & = & \frac{1}{2} R_{\rm mag} H_{\rm RF}^2 \\
R_{\rm mag} & = & \frac{\omega^2}{(\omega_0-M\omega^2/\eta)^2 + (\omega)^2} \frac{B_{\rm res}}{2B_{\rm c2}} R_{\rm n}. \label{eq:old_model_derived}
\end{eqnarray}
The effective inertia of a flux $M$ is usually negligible.
In a flux flow regime $\omega \gg \omega_0$, Eq.~(\ref{eq:old_model_derived}) reproduces Eq.~(\ref{eq:old_model}).

In a flux pinning regime $\omega < \omega_0$ and an intermediate condition $\omega\sim\omega_0$,
the flux oscillation contains an additional factor
\begin{equation}
\frac{\omega^2}{(\omega_0-M\omega^2/\eta)^2 + (\omega)^2}
\end{equation}
that depends on material parameters and RF frequency.
In our study, frequency is relatively low ($<352$~MHz) and the cavities are made of a clean polycrystalline niobium.
It is not possible to precisely know the relation between $\omega$ and $\omega_0$.
The good agreement of Eq.~(\ref{eq:new_model}) with a geometrical convolution Eq.~(\ref{eq:new_model}) may indicate that our cavities are in the flux flow regime.

\section{Quench and flux trapping\label{sec:quench}}
Quenching a cavity during operation could potentially lead to the degradation of its quality factor due to fast flux entry into a normal conducting quench spot~\cite{laxdal13}. 
This degradation is fully extrinsic and it only depends on the external residual magnetic field around the cavity during quench~\cite{PhysRevApplied.5.044019}.
Also, it is of importance to study the evolution of a quench spot by using the dynamics of flux penetration followed by flux rearrangement. 
The sensitivity of the fluxgate sensor enables us to address these phenomena.

\begin{figure}[h]
\includegraphics[width=85mm]{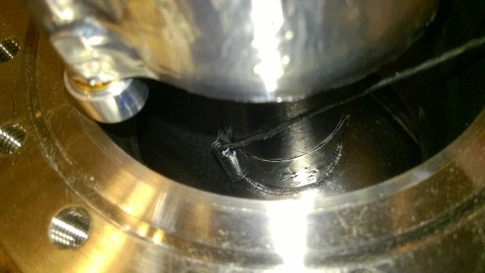}
\caption{
Magnetic sensor at the quench location (bottom of the port) and second sound transducer (top of the port) installed on a Spiral2 prototype QWR.
\label{fig:quench_localized}}
\end{figure}

So as to study this on Spiral2 QWR, we installed a magnetic sensor probing the vertical magnetic component at the quench location as shown in Fig.~\ref{fig:quench_localized} and also indicated in Fig.~\ref{fig:cavities} as a probe named quench. 
Before this experiment, previous works~\cite{Longuevergne:2015ass, Fouaidy_2017} had localized the quench location. 
In this particular experiment, only one of the fluxgate sensors is read out without multiplexing in order to catch the fast quench events. 
For signal amplitudes less than 10~mG, the response time of the fluxgate sensor is faster than 30~ms~\cite{fluxgate_manual} while the data acquisition rate limits the time resolution to 50~ms.

\begin{figure}[h]
\includegraphics[width=85mm]{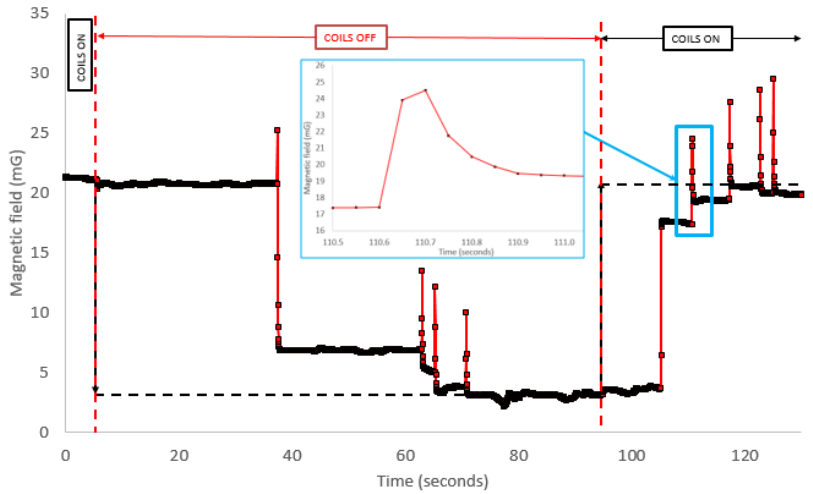}
\caption{
Magnetic flux trapping during cavity quench.
\label{fig:quench}}
\end{figure}

Figure~\ref{fig:quench} depicts how the magnetic field at the quench location is changing after several quenches. 
Before quenching the cavity, the compensating coils are switched off at time 5~s to change the magnetic background as indicated by the dashed black line. 
As presented in the main text, the inner conductor completely shields this magnetic field and results in no change of the measured field by the probe. 
After the first quench around time 38~s, the field promptly drops significantly, indicating the flux penetration into the quench spot. 
Such a flux entry is possible because of the weak shielding provided by the outer conductor. 
The measured magnetic field reaches saturation around 3 mG after several quenches around time 70~s. 
After the saturation, the coils are switched back on and the field is compensated again at time 95~s. 
Then, after a couple of quenches, the measured magnetic field saturated back to the initial level around time 130~s. 
The cavity has to undergo at least three quenches to reach saturation

The transient response of the magnetic field to the quench event is characterized by a narrow peak of width 50~ms (standard deviation),
followed by an exponential relaxation with time constant 75 ms and is eventually stabilized to a constant floor as shown in the inset plot in Fig.~\ref{fig:quench}. 

The observed peak is due to the demagnetization effect at the opening of a normal conducting quenched area, 
in which the magnetic field contained between the outer and inner conductor can tunnel. 
According to our simulation, the demagnetization factor becomes the maximum when the radius of the quench spot is around 10~mm. 
Our previous study~\cite{Lesrel:1997rb} showed that the time scale of hot spot expansion is less than 1~ms; 
therefore, this phenomenon is smeared by the time resolution of the detector response. 

On the other hand, the quench zone cools down with a characteristic time of the order of 50~ms~\cite{Lesrel:1997rb} and eventually collapses to be superconducting again. 
The flux lines penetrating the wall during the quench are pushed inwards by the phase front. 
As we discussed in the main text, almost all the flux would be trapped by pinning centers of the polycrystalline material to relax the demagnetization at the phase front. 
Therefore, opening and closing the quenched spot is an irreversible process and results in exponential relaxation in the measurement.

Finally, the trapped flux is frozen and results in the constant floor. 
This determines the total number of trapped flux during a single quench event. 
Multiple quenches let magnetic flux quanta occupy all the potential pinning centers, similar to an observation in an elliptical cavity [7]. 

In conclusion of this section, flux trapping happens during quench and could be reversible. 
Several cycles of quench are necessary to reach saturation. 
It is thus possible to recover a  degradation triggered by a quench event by re-quenching the cavity in a re-optimized magnetic environment instead of warming up a full cryomodule above transition as reported in~\cite{PhysRevApplied.5.044019}. 
The  degradation by quench (if caused by flux trapping) is caused by a non-optimal magnetic shielding.

\section{Surface barrier and the flux orientation}\label{sec:image_force}
When the fluxes are carefully trapped on purpose, our hypothesis of preferential flux trapping normal to the surface can reproduce experimental results reported in~\cite{Kramer:2019nzd}. 
However, their magnetic field measurement outside the cavity apparently showed good agreement with a homogeneous trapping scenario without any preferences in flux orientation. 
In this section, we consider that the flux can be trapped in parallel to the surface but does not contribute to the power dissipation which happens at the inner surface.

\begin{figure}[h]
\includegraphics[width=85mm]{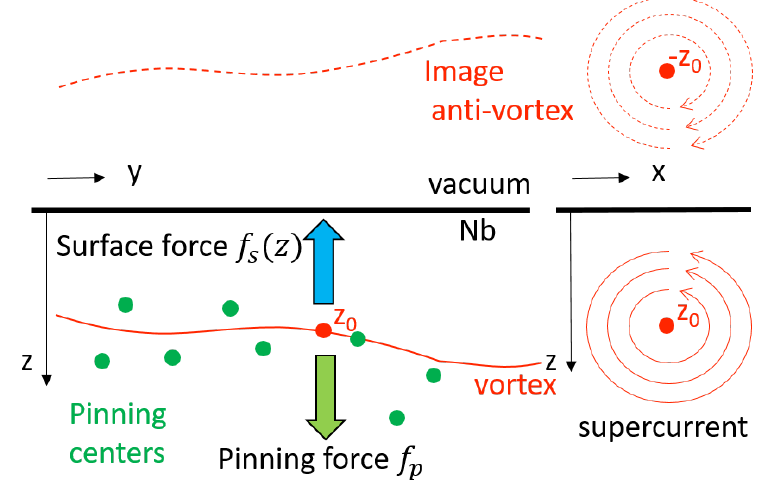}
\caption{
Flux line trapped in parallel to the surface.
\label{fig:Bean}}
\end{figure}

We apply the surface barrier model by Bean and Livingston~\cite{PhysRevLett.12.14} to our case. 
We first assume that our cavity is a local superconductor but is still relatively clean, 
because we do not perform any low temperature baking and/or nitrogen doping and infusion. 
The trapped flux is also assumed to be sufficiently smooth. 
The magnetic field of one vortex trapped parallel to the surface can be obtained by the modified London equation\cite{vallet1994, matsushita2014}
\begin{eqnarray}
&& \nabla^2 H(x, z)-\frac{1}{\lambda^2}H(x, z) \nonumber \\
&=& -\frac{\phi_0}{\mu_0 \lambda^2}\left[ \delta(x)\delta(z-z_0) -  \delta(x)\delta(z+z_0)\right], \label{eq:Bean}
\end{eqnarray}
where $\lambda$ is the London penetration depth and $z_0$ is the depth of the flux. 
We take the coordinate system as shown in Fig.~\ref{fig:Bean}. 
The image force method is used to fulfil the boundary condition at the surface. 
If such a vortex exists within a few penetration depths, where the RF current is along the x-axis, 
it vibrates in the z-direction and contributes to additional power dissipation.

The solution of Eq.~(\ref{eq:Bean}) is a sum of one particular solution of the Green function of the two-dimensional inhomogeneous Helmholtz equation and a general solution of the homogeneous equation without the source term. 
The former gives the image force per unit length, {\it attractive} to the surface
\begin{equation}
f_1(z_0) = \frac{\phi_0}{2\pi\mu_0\lambda^3} K_1\left( \frac{2z_0}{\lambda}\right)
\end{equation}
with $K_1$ the modified Bessel function of the second kind. 
The latter is a {\it repulsive} force per unit length from the interaction to the external magnetic field
\begin{equation}
f_2(z_0) = \frac{\phi_0H_0}{\lambda} \exp{\left(-\frac{z_0}{\lambda} \right)}
\end{equation}
with a constant $H_0$ which satisfies continuity of the magnetic field at the interface between niobium and the vacuum. 

Bean and Livingston determined the surface barrier by relating these counter-acting forces: $f_{\rm s}(z_0) = f_1(z_0)-f_2(z_0)$. 
As is well known, even a small $H_0$ generates a finite surface barrier, 
which prevents a trapped flux from escaping toward the vacuum. 
In our configuration, a very small $H_0$ of maximum 100~mG results in the peak of the surface barrier i.e. $f_{\rm s} = 0$ deep inside the bulk. 

We compare~\cite{vallet1994} the image force to another force from the pinning centers. 
This pinning force further prevents the escaping trapped flux as a frictional force. 
In general, estimating the pinning effect is very difficult~\cite{PhysRev.178.657} and its strength can vary by several orders of magnitudes with impurity contents, dislocations, precipitates, grain boundaries etc. 
Here, we evaluate this in two ways. 
In the following discussion, we take the parameters of clean niobium: 
lower critical field $\mu_0H_{\rm c1}(0)=170$~mT, 
thermodynamic critical field $\mu_0H_{\rm c}=200$~mT, 
upper critical field $\mu_0H_{\rm c2}=410$~mT, 
coherence length $\xi=39$~nm,
penetration depth $\lambda=32$~nm~\cite{padamsee1998, saito2003, CASALBUONI200545, Dhavale_2012}.

First, we estimate the lower bound of the pinning force using our results. 
The almost full trapping of the flux during cooling down was observed in our cavities. 
This implies that the pinning force is at least stronger than the thermal force~\cite{PhysRevApplied.5.044019, PhysRev.181.701}
\begin{equation}
f_{\rm T} = S\Delta T < f_{\rm p}
\end{equation}
with the transport entropy $S$ per unit length
\begin{equation}
S = -\phi_0 \frac{\partial H_{\rm c1}}{\partial T}.
\end{equation}
If we take an empirical formula 
\begin{equation}
H_{\rm c1}(T)\sim H_{\rm c1}(0)\left[1-(T/T_{\rm c})^2 \right],
\end{equation}
we get
\begin{equation}
S = 2\phi_0 H_{\rm c1}(0)\frac{T}{T_{\rm c}^2}\sim 5.6\times10^{-10} \frac{T}{T_{\rm c}^2}.
\end{equation}
The maximum $\Delta T$ was 80~K between a typical cavity size of 1~m at transition $T\sim T_{\rm c}=9.25$~K, and we estimate
\begin{equation}
f_{\rm p} > f_{\rm T}\sim 4.8\times 10^{-9}\, {\rm N/m}. \label{eq:fp_from_thermal}
\end{equation}
Since the cavity is cooled down along the wall, this force is in either x or y directions in Fig.~\ref{eq:Bean}. 
We assume isotropic pinning effects and apply this lower bound also to the z-direction along which the parallel fluxes migrate.

Next, we estimate the pinning force from the sample experiments on critical de-pinning current density $J_{\rm c}$, 
at which trapped vortex starts to escape from the pinning centers by Lorentz force; thus,
\begin{equation}
f_{\rm p} = \left|\mathbold{J}_{\rm c} \times \mathbold{\phi}_0 \right|.
\end{equation}
On low-purity niobium, Das Gupta et al. obtained~\cite{doi:10.1063/1.322862} $J_{\rm c}\sim5\times10^10$~A/m$^2$ near the surface and $J_{\rm c}\sim2\times10^9$~A/m$^2$ at 3~$\mu$m deep inside the bulk.
Recently, a more relevant experiment on clean fine grain niobium for the cavity application showed~\cite{Dhavale_2012} $J_{\rm c}=10^8$-$10^9$~A/m$^2$ and therefore we can estimate
\begin{equation}
f_{\rm p} = (2\times10^{-7} - 2\times10^{-6}) \, {\rm N/m}.
\end{equation}
The large uncertainty of a factor 10 comes from different models to estimate $J_{\rm c}$ from DC magnetization. 
These results are consistent with the estimation Eq.~(\ref{eq:fp_from_thermal}).
Figure~\ref{fig:image_force} compares the estimated pinning force and the surface forces.

\bibliographystyle{apdrev}
\bibliography{david}
\end{document}